\documentclass{article}

\usepackage{arxiv}

\usepackage[utf8]{inputenc} 
\usepackage[T1]{fontenc}    
\usepackage[colorlinks=true,linkcolor=blue, citecolor=blue, urlcolor=blue]{hyperref}       
\usepackage{url}            
\usepackage{booktabs}       
\usepackage{amsfonts}       
\usepackage{nicefrac}       
\usepackage{microtype}      
\usepackage{graphicx}
\usepackage{lmodern,textcomp}
\usepackage{subcaption}
\usepackage{amsmath}
\usepackage{amsbsy}
\usepackage{longtable}
\usepackage{lscape}
\usepackage{verbatim}
\usepackage{float}
\usepackage{booktabs}
\usepackage{dcolumn}
\usepackage{comment}
\usepackage{multirow}
\usepackage{amsthm,amsmath}
\usepackage{tabularx}
\usepackage[section]{placeins}
\graphicspath{ {./images/} }

\title{Gender bias in the Erasmus students network}

\author{
 Luca De Benedictis \\
  University of Macerata and Luiss\\
  \texttt{luca.debenedictis@unimc.it} \\
   \And
Silvia Leoni \\
  Marche Polytechnic University\\
  \texttt{s.leoni@univpm.it} \\
}

\begin{document}
\maketitle
\begin{abstract}
The Erasmus Program (EuRopean community Action Scheme for the Mobility of University Students), the most important student exchange program in the world, financed by the European Union and started in 1987, is characterized by a strong gender bias. Girls participate to the program more than boys. This work quantifies the gender bias in the Erasmus program between 2008 and 2013, using novel data at the university level. It describes the structure of the program in great details, carrying out the analysis across fields of study, and identifies key universities as senders and receivers. In addition, it tests the difference in the degree distribution of the Erasmus network along time and between genders, giving evidence of a greater density in the female Erasmus network with respect to the one of the male Erasmus network. 
\end{abstract}



\section*{Introduction and prior research on the Erasmus program}
At its 30\textsuperscript{th} anniversary celebrations in 2017, the Erasmus program counted more than 9 million participants since its creation, increased to more than 10 million in 2018.
The program, which allows its participants to study or take an internship in a foreign country,\footnote{The Erasmus program includes also a variety of other initiatives, such as volunteering activities, Erasmus Mundus Joint Master Degrees, the European Week of Sports, Jean Monnet Modules and Chairs.\\} has become very popular among university students whose participation is increasing year after year. Its popularity has made it a true cultural phenomenon, 
and, given the successful outcome of Erasmus+, the European Commission has proposed, for the 2021-2027 plan, to double the funds allocated to the program in order to support the mobility of 12 million people, making the program more inclusive, allowing the participation of those coming from disadvantaged families background and less inclined to international mobility. With its objective of inclusion, the program has also become a cornerstone for equal opportunities, with many of its projects, for example, directly aimed at the promotion of gender equality. Nevertheless, the participation in the Erasmus program itself is characterized by a remarkable and long-lasting gender bias, that favours women, since its launch. 

Evidence shows that the number of women participating in the program has been higher than that of men up to the 1990s \cite{maiworm2001erasmus}. This gender gap is reported by \cite{bottcher2016gender} for the academic year 2011-12, across both countries and subject area. In this cross-sectional analysis, the authors specifically study the over-representation of female Erasmus students by comparing tertiary education statistics with the Erasmus data using a null model for which it is assumed that the population of Erasmus students is randomly drawn from the student population of participating countries. 
More broadly, the evidence of a bias in favour of female students in the Erasmus program can be related to the research suggesting that women participation in global academic mobility is large due to the growing number of women enrolled in higher education, as well as the greater gender parity across the world \cite{bhandari2017women}. Globally, this pro-female trend may be attributed also to the opportunities provided by targeted scholarship and fellowship programs for under-represented groups to pursue advanced study outside their home countries. The development of emerging economies and women's emancipation in developing countries also play a role, as in the case of China, which saw an economic boom along with a change in the role of women, who became more individually-oriented and delayed their marriage age \cite{martin2014gender}. Nevertheless, under-representation of women among inbound international higher education students is highly significant in some countries, as in the United States, where although the progress made in the last decades, the gender imbalance in terms of incoming international students is still widening, possibly because of the rising number of international students in STEM-related fields which are typically dominated by male students and, at the same time, very attractive to international students in the United States \cite{bhandari2017women,myers2019geography}. In Europe, the United Kingdom, instead, hosts more female international students rather than male students \cite{myers2019geography} and also shows evidence that U.K. female graduates are more mobile than men graduates \cite{faggian2007some}. The authors rationalize this outcome by suggesting that women use migration as a form of compensation for gender discrimination in accessing the local labour market since spatial employment exploration could allow identifying better job opportunities. 

This work solely focuses on Erasmus mobility for study reason. The aim is to analyse the gender difference in the participation in the program and the elements contributing to the existence and persistence of this bias.
We use the data available on the EU Open Data Portal, which correspond to six datasets, describing students flows between universities in European countries since a.y. 2008/09 until 2013/14.

First of all, the paper explores the Erasmus students network and studies the relationships between universities in the network. By studying the weighted directed network of students the paper identifies the most important hubs hosting and sending students abroad, providing a picture of the participation to the program, by gender and across fields of study.
A network approach to the Erasmus mobility is also used in \cite{derzsi2011topology}, which analyse Erasmus students mobility in 2003 by studying the related non-directed and non-weighted graph and its directed and weighted graph, where nodes are represented by European universities and links are connections between pair of universities. 
The authors focus on the study of network properties and topology and their findings suggest that the degree distribution of the network follows an exponential distributional model and that the Erasmus network of universities cannot be considered a scale-free network, but rather a small-word type of network with a giant component.

Secondly, gender imbalance is further studied in terms of the degree distribution of the Erasmus network. Considering the directed and unweighted Erasmus network of universities, the analysis explores the possible changes in the indegree and outdegree distributions along time and between gender and tests a power law fitted model to the data.

The paper is organized as follows: the first section provides a description of the data used in the analysis; the second section performs a quantification of the Erasmus program general trend and gender imbalance in student flows across fields of study; the third section analyses the networks that can be identified in the data; the fourth section compares the network indegree and outdegree distributions over time and between genders; finally, the conclusive section summarizes the results of the analysis and draws some final observations regarding the possible future evolution of the gender bias in the Erasmus program. 

\section*{Data}
Data used in the analysis are available at the EU open data portal and are freely accessible. They consist of 6 datasets corresponding to the academic years from 2008/09 to 2013/14 and they contain observations for each participant to the mobility in the relative a.y. including information on the type of mobility (study or placement), the home country and the host country, the home university and the host university, the field of study coded in the ISCED 1997-2011 classification \cite{uis2012international}, the participant gender, the level of study (first or second cycle), the duration of the mobility, the amount of the grant received and the language used in the mobility. We only consider observations related to study mobility type and clean the data of mistakes present in the original datasets. 

By counting the frequencies of unique connections between pairs of universities, we specify the weight of each connection, that represents the students flow, i.e. the number of students going on mobility from university A to university B. Data are therefore collapsed so to have observations for each dyad of universities linked by at least one student in mobility, by gender and by field of study.
The fields of study are grouped in macro fields according to the ISCED-F 2013 coding system \cite{unesco2014isced}, which identifies 12 macro fields. 

Eventually, for the whole period of time considered, the datasets count 762304 observations in total, with 3148 universities present in the time span considered.

By providing source and target identity for each observation, such data format easily allows to identify a network structure within the data. In particular, information at university level allows to build the Erasmus network of universities, offering a very detailed level of analysis, which so far has been little explored in the literature (an example is found in \cite{derzsi2011topology}). Data allow also to consider both the unweighted and the weighted network, when accounting for the flow of students moving from one university to another one abroad. In addition, knowing the participants' gender permits to study the gender balance in the students flows, as well as to decompose the unweighted network in the male-network and the female-network, in order to investigate potential gender imbalances in the connections between universities. Finally, the analysis can be conducted across field of study and considering a significant time span, with respect to the cited literature.

\section*{The Erasmus program}
\label{sec:headings}
Erasmus stands for European Region Action Scheme for the Mobility of University.\footnote{The name of the program comes from the Latinisation of the name of Erasmus of Rotterdam. The success of this program has led to the use of naming other European programs after famous personalities from various European cultures such as Socrates, Leonardo or Comenius \cite{corradi2015}.} It is a student mobility program created by the European Union in 1987. The program started with the idea of allowing European university students to study abroad in a European university, with the legal recognition of the mobility in the home university and providing a scholarship to cover the additional cost for studying in another country of the EU for a period of between three months and one year.
The work that led to the official approval of the program saw the involvement of universities from all over Europe in order to establish the legal and financial basis necessary to develop and manage organisational and educational cooperation between universities underpinning the Erasmus program \cite{corradi2015}.

The objectives of Erasmus+ and the original Erasmus program can be summarised in strengthening the European identity, increasing individual skills and, thus, their employability.
By creating opportunities for study, training, work experience and volunteering abroad, Erasmus aims to respond to the problems of unemployment and skills shortages in Europe and to modernise education and training systems.
Erasmus+ is, in fact, part of a socio-economic context that, on the one hand, counts more than 3 million young Europeans unemployed in 2019, with youth unemployment rates reaching 30\% in some countries, and on the other hand, almost 4 million job vacancies, with more than a third of employers reporting difficulties in recruiting staff with the required qualifications\footnote{Source: Eurostat 2019.}. In addition, the Erasmus program is considered a successful example of European integration and a symbol of the construction of European identity.

\subsection*{General trend}
\begin{figure}[h!]
  \centering
  \includegraphics[scale=0.45]{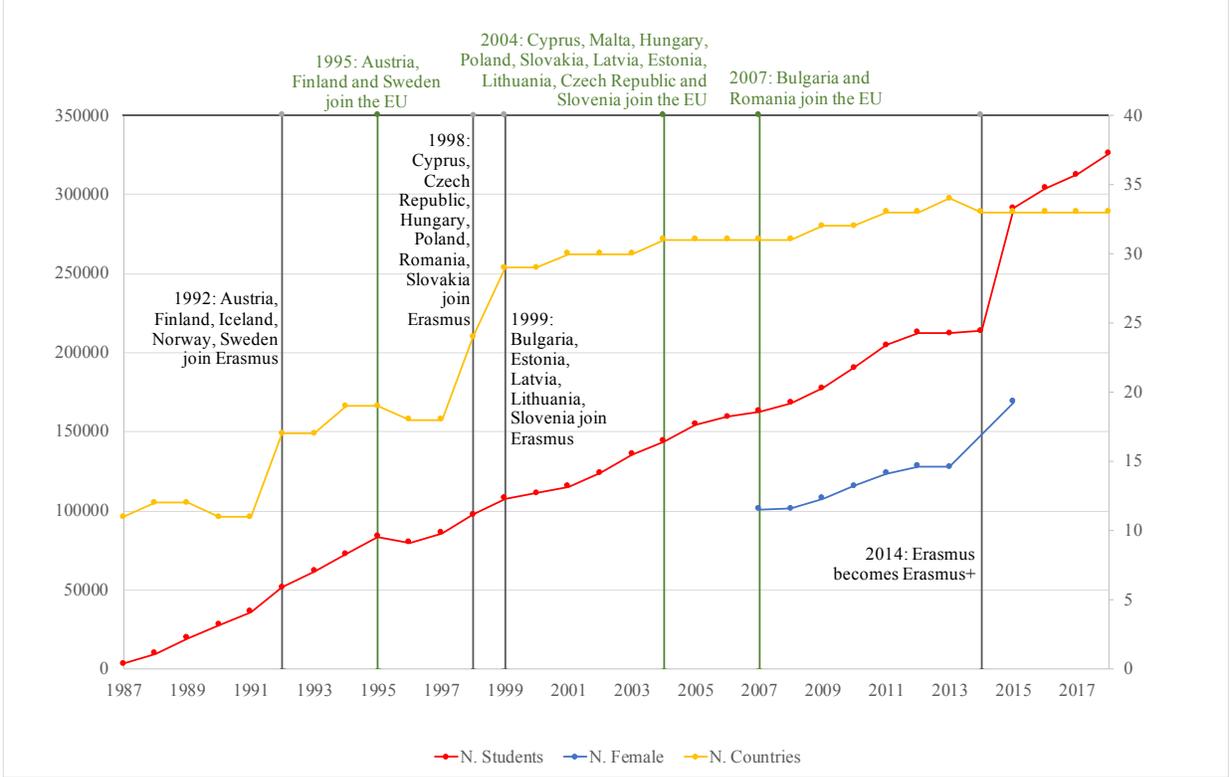}
  \caption{The history of Erasmus program from 1987 to 2018.
       The evolution of the Erasmus program is represented in terms of countries and students on mobility only for study reason. The red line represents the number of mobile university students (left scale), the blue line represents the number of female university students on mobility only for the years in which the information is available (left scale), the yellow line represents the number of countries participating in the mobility (right scale). The grey vertical lines indicate the adhesion of new countries to the Erasmus program and the transformation of the program into Erasmus+. The green vertical lines identify some enlargements of the EU. Note that the value for the year 2014 is not an \textit{ex post} figure, but a projection \cite{european2015erasmus+}.}
      \label{fig:fig1}
      \end{figure}

Over time, Erasmus has become an essential part of the unified European mobility programs in the area of education Socrates I (1994-1999), Socrates II (2000-2006) and Lifelong Learning Program (2007-2013) and has grown in size.
The program started in 1987 with 11 participating countries and 3244 students on mobility and reached 33 participating countries in 2018 and 325495 university students on mobility for study reason (see Figure \ref{fig:fig1}). 

The growth in the number of participating students has followed the growing trend of the countries participating in the program, which in many cases joined Erasmus before becoming EU Member States.
This is the case of Austria, Finland and Sweden, for example, which became EU Member States in 1995 (as shown in Figure \ref{fig:fig1}), but joined the Erasmus program as early as 1992; or the case of those countries that joined the EU in 2004 but have been participating in Erasmus since 1998 and 1999. The number of students participating in the mobility saw an unprecedented increase in 2015, going from 213879 to 291383, although the number of participating countries remained almost unchanged. In 2014 the program became Erasmus+ and changed its structure (with the EU Regulation 1288/2013): it is no longer exclusively dedicated to education, but also to training, youth and sport, and it no longer restricts participation to university students only, but also admits, for example, school and university teaching staff, as well as administrative staff. 
Therefore, it is an integrated program which has incorporated all the funding mechanisms for school and university student mobility implemented by the European Union until 2013 (e.g. Comenius, Leonardo Da Vinci and others). In 2013 the program reaches the largest number of participants, 34, the 28 EU Member States plus Switzerland, Iceland, Liechtenstein, Norway, North Macedonia and Turkey. Since 2014, Switzerland no longer enjoys the status of participant to the program, but it is now a partner country, i.e. it has adopted a transitional solution financed with Swiss funds which still allows Swiss people and institutions to take part in the program. In 2019, participant countries have been 34 again with the official entry of Serbia.
€14.7 billion were allocated to the Erasmus budget for the period 2014-2020, $40\%$ more than the previous programming period, and, as already highlighted, for the period 2021-2027 the European Commission has proposed to double the figure to €30 billion \cite{europeancommission2018}. 
The program continues to grow with the aim of becoming more powerful and inclusive.

\subsection*{Gender balance}

\begin{figure}[h!]
\centering
\includegraphics[scale=0.5]{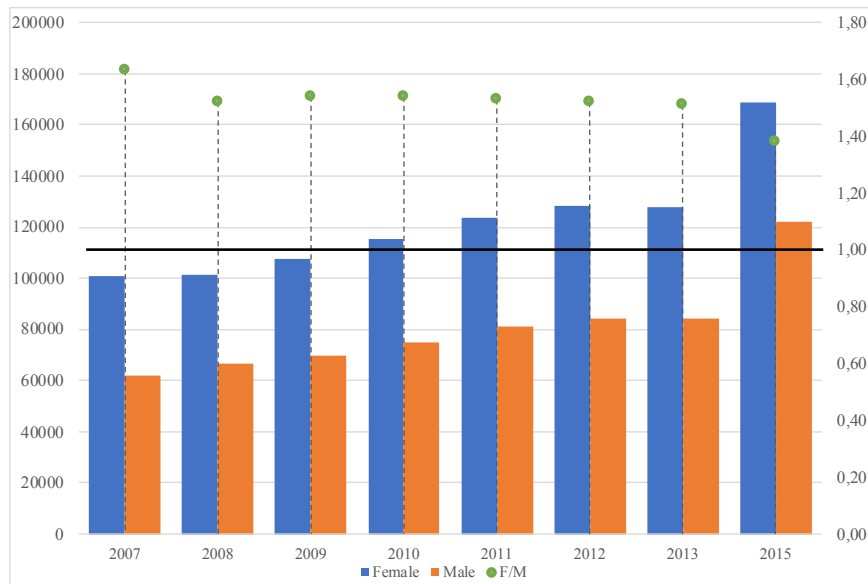}
	\caption{Gender balance in the Erasmus mobility over selected years.
		 The blue bars indicate the number of female Erasmus students, the orange bars indicate the number of male Erasmus students (both to be read on the left vertical axis). The green indicator represents the value of the female over male ($F/M$) ratio (right vertical axis). The horizontal line positioned at the ordinate value of 1 on the right vertical axis corresponds to $F/M=1$.}
	 \label{fig:fig2}
\end{figure}

Figure \ref{fig:fig2} shows the amount of female and male participants in the Erasmus program for some selected years and the relative ratio between female and male students. The number of both female and male students grow following the growing general trend in participation seen in Figure \ref{fig:fig1}, with female representing a large majority for each year. Nevertheless, the ratio $F/M$ slightly decreases over the years considered signaling a wider participation of men. 

\begin{figure}[h!]
\centering
\includegraphics[scale=0.45]{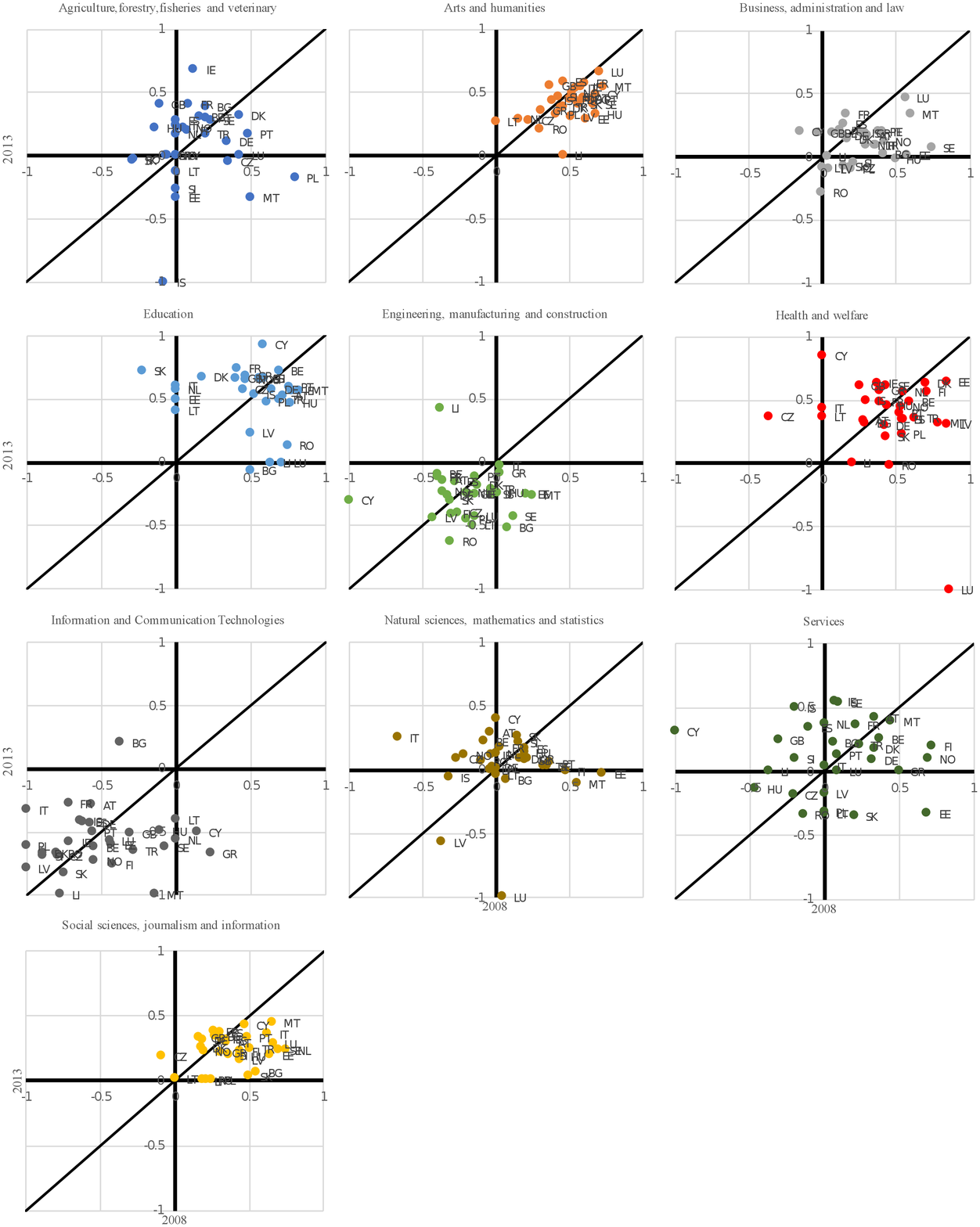}
	\caption{Gender balance in the incoming flows across fields of study for the year 2008 against 2013.
		The scatter plots show the $F/M^{B}$ measure for the incoming flows of students for the years 2008 and 2013 for ten fields of study corresponding to the ISCED-F 2013 classification. Two macro fields (\textit{Not known or unspecified} and \textit{Generic programs and qualifications}) are not showed.}
	 \label{fig:fig3}
\end{figure}

\begin{figure}[h!]
\centering
\includegraphics[scale=0.45]{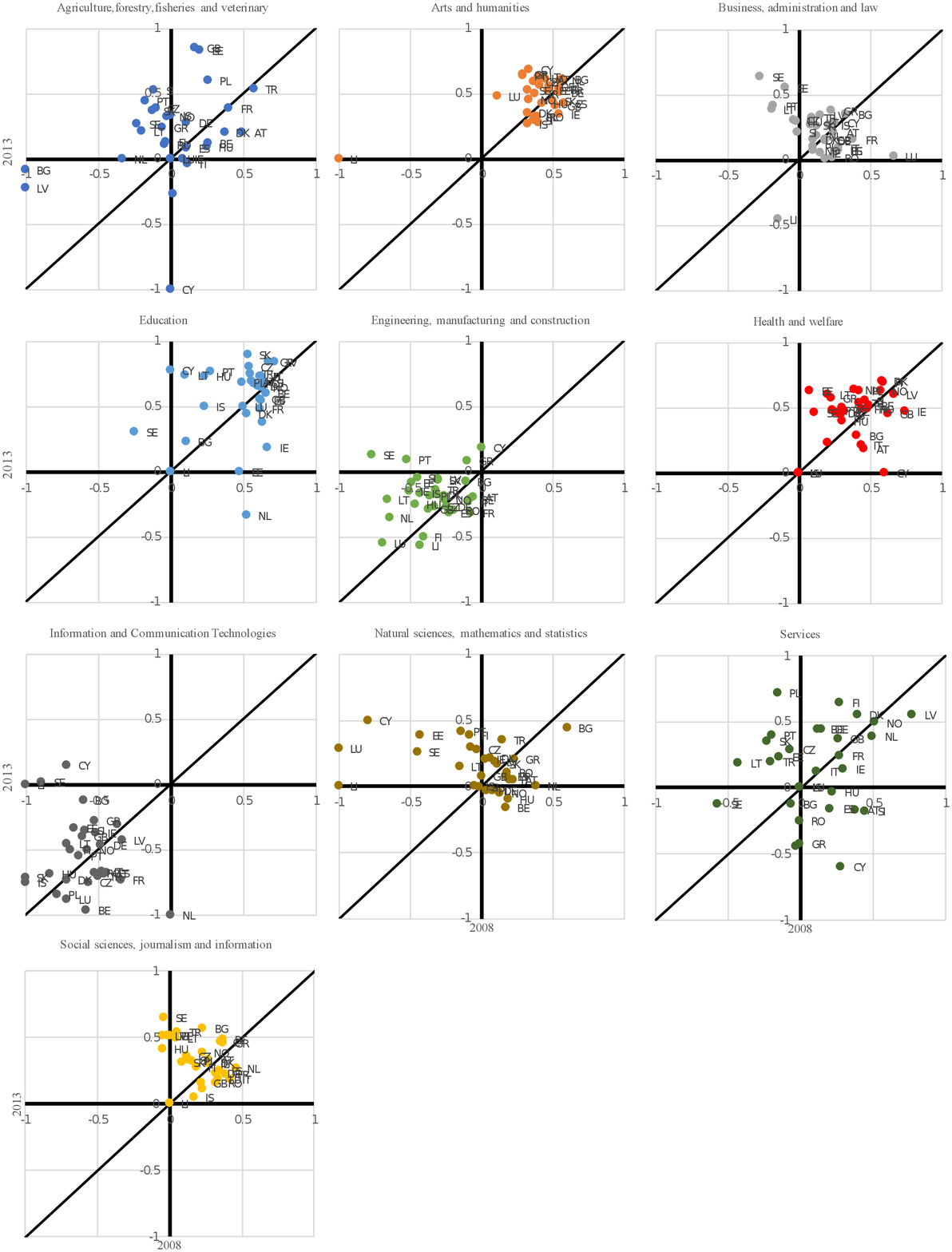}
	\caption{Gender balance in the outgoing flows across fields of study for the year 2008 against 2013.
		The scatter plots show the $F/M^{B}$ measure for the outgoing flows of students for the years 2008 and 2013 for ten fields of study corresponding to the ISCED-F 2013 classification. Two macro fields (\textit{Not known or unspecified} and \textit{Generic programs and qualifications}) are not showed.}
	 \label{fig:fig4}
\end{figure}

Differences between gender can be observed more in detail across fields of study. In this case, to obtain a clearer visualization, we use a transformation of the ratio $F/M$ based on \cite{benedictis2005three}, given by:
\[
F/M^{B} = \frac{(F/M) - 1}{(F/M)+1},
\]
where the superscript $B$ stands for \textit{bounded}. The $F/M^{B}$ index provides a measure of the comparative advantage of female participation over male participation (or viceversa) with a value ranging (\textit{bounded}) between $[-1,1]$ and demarcation value equal to 0, corresponding to absence of bias. 
Figures \ref{fig:fig3} and \ref{fig:fig4} plot this measure for 2008 against 2013 for each macro field of study, respectively for incoming and outgoing students.

Some fields of study show a predominance of female over male students which remains pretty stable over the time span considered, as in the case of Art and humanities and Education, where a cloud of countries is concentrated around the bisector of the first quarter, with the exception, for the Education sector, of some Eastern countries (see Bulgaria, Romania and Latvia for instance) in the incoming flows (Figure \ref{fig:fig3}), which show a decrease of the female predominance from 2008 to 2013, and in the outgoing flows of the extreme cases of the Netherlands and Sweden which had a reverse of the index respectively in favor of men and women (Figure \ref{fig:fig4}). 

A somehow specular situation can be identified in the field of Engineering, manufacturing and construction where the male predominance characterizes the sector, although the cloud is less concentrated around the bisector and rather flattened, taking for the incoming flows a horizontal shape, which pushes towards the right side, revealing that for a few countries such as Bulgaria, Sweden, Estonia and Hungary the sector once dominated by female students in mobility changed tendency with a predominance of male students in 2013 (Figure \ref{fig:fig3}). For the outgoing flows the cloud of countries shows more stability over time, with the exception of Sweden, Portugal and Greece, where the field once heavily dominated by outgoing male students had a reversed tendency in favor of female students in 2013 (Figure \ref{fig:fig4}).

A horizontal cloud shape can be identified also in the field of Health and welfare, for example, both for the incoming and the outgoing flows. Especially for the former, a group of Eastern countries (see Latvia, Poland, Slovakia) saw a decrease of the measure of bias thus in favor of a growing number of male students, up to cancelling out the bias in the case of Romania.

The same can be observed in the incoming flows of students in Business, administration and law, Social sciences, journalism and information, where about the same group of Eastern/Northern countries reduces the gender bias over time in favor of a greater male participation, and in the Information and Communication Technologies (ICTs), where Mediterranean countries see a tendency towards gender parity in a field for the most part dominated by male students. 

On the contrary, note the vertical shape that the cloud of countries take for the outgoing flows in the ICTs and Social sciences sectors, revealing an increase in gender disparity over the years considered. 

Finally, in the fields of Agriculture, forestry, fisheries, veterinary and Services it is not possible to identify a general trend, as the cloud of points show heterogeneity in the gender balance among countries; in the field of Natural sciences, Mathematics and Statistics, instead, the cloud of countries is concentrated in proximity of the origin of the axis, pointing out gender equity along time, in a sector usually characterized by a larger male presence as common in the STEM-fields \cite{botella2019gender,oecd2017pursuit}.

Briefly, decomposing the analysis over time across fields of study confirms what observed in general in Figure \ref{fig:fig2}. The gender imbalance in participation in favor of women has been decreasing over the time span considered and, in particular, this reduction is observed for the groups of Eastern and Mediterranean countries.

\clearpage 

\section*{Plain Vanilla Network Analysis}
\label{sec:vanilla}

In this section we will visualize, describe and summarize the characteristics of the Erasmus program, comparing its structure in 2008, the initial year in our time span, with 2013, the last year, keeping the resulting network of student flows separated by gender.

The Erasmus Network ($\mathcal{N}$) is defined by the graph $\mathcal{G}_{t}=(\mathcal{V}_{t}, \mathcal{L}_{t})$, the edge value function, $\mathcal{W}$, and the node value function $\mathcal{O}$, for every year $t \in [2008, 2013]$. In case of $t$=2008,  $\mathcal{G}_{2008}=(\mathcal{V}_{2008}, \mathcal{L}_{2008})=(3148, 102310)$; the graph of the female flows is $\mathcal{G}_{2008}^F=(\mathcal{V}_{2008}^F, \mathcal{L}_{2008}^F) = (3148, 59384)$, while the one of men is $\mathcal{G}_{2008}^M = (3148, 42926)$. The simple comparison of the dynamics of $\mathcal{L}^F$ and $\mathcal{L}^M$ makes the first characteristic of the gender bias in the Erasmus Network evident: the ratio between $\mathcal{L}^F$ and $\mathcal{L}^M$ was 1.383 in 2008 ($\mathcal{L}_{2008}^F$ were 58.0\% of total links, $\mathcal{L}_{2008}$, and $\mathcal{L}_{2008}^M$ were 72.285\% of $\mathcal{L}_{2008}^F$) and in 2013 it was 1.387 in 2013 ($\mathcal{L}_{2013}^F$ were  58.1\% of $\mathcal{L}_{2013}$, and $\mathcal{L}_{2013}^M$ were 72.069\% of $\mathcal{L}_{2013}^F$. The values oscillate during the years but the gender bias remains quite persistent. The same is even more evident if we observe it though a density decomposition.

In a directed network, like the one considered here, density is formally defined as $\gamma = \frac{m}{m_\text{max}} \equiv \frac{m}{n (n-1)}$,
where $m$ is the number of observed arcs in $\mathcal{N}$, while $n$ is the number of nodes.  
Among the many properties of the density, two are particularly handy, in the present context: the first one is that being a measure bounded between 0 and 1 it can be interpreted in a probabilistic way; the second one is that it can be decomposed among the different contributions of $\mathcal{L}^F$ and $\mathcal{L}$ to the overall density. 

The overall density $\gamma$ is rising from a value of 0.010 to 0.013, with a decreasing slope: remaining a quite sparse network, $\mathcal{N}$ is becoming more connected, and if the probability, for two universities chosen at random, to share an Erasmus program was 1.0\% in 2008, it was 1.3\% in 2013.\footnote{ The formula of the density is in this case $\gamma_{2008} = \frac{m_{2008}}{n (n-1)} = \frac{102310}{3148 \times 3147} \equiv \frac{59384}{3148 \times 3147} + \frac{42926}{3148 \times 3147} \equiv 0.01 \equiv 0.006 + 0.004  \equiv \gamma_{2008}^F + \gamma_{2008}^M$. In 2013, $\gamma_{2013} \equiv \gamma_{2013}^F + \gamma_{2013}^M \equiv 0.008 + 0.005$.}

The evolution of $\mathcal{N}$ from 2008 to 2013, made 340 (among the 939) isolated links in $\mathcal{G}_{2008}^F$ and 327 (among the 1031) isolated links in $\mathcal{G}_{2008}^M$ attached to the connected component of $\mathcal{N}_{2013}^F$ and $\mathcal{N}_{2013}^M$, respectively. The heterogeneity in connectivity is relevant and multiform: (1) the node with the highest \emph{indegree} centrality, $v_{max}$, had a $\text{deg}(v_{max})$ equal to 669 for $\mathcal{G}_{2008}^F$ (it was the Universidad de Granada, in Spain), and to 403 for $\mathcal{G}_{2008}^M$ (the Universitat Politecnica de Valencia, in Span); it become equal to 668 and 392 for $\mathcal{G}_{2013}^F$ and $\mathcal{G}_{2013}^M$, with the Universidad de Granada reaching the top position in both networks. (2) As far as \emph{outdegree} centrality, the $\text{deg}(v_{max})$ was equal to 537 for $\mathcal{G}_{2008}^F$ (it was the Universidad Complutense de Madrid, in Spain), and to 541 for $\mathcal{G}_{2008}^M$ (the Univerzita Karlova of Prague, in the Czech Republic) and reached the value of 622 and 484 for $\mathcal{G}_{2013}^F$ and $\mathcal{G}_{2013}^M$, with the Universidad de Granada on top of both ranks.

Being so sparse and also showing some major hubs and authorities, $\mathcal{N}$ is all cases not characterized by a hierarchical structure. Fitting $\mathcal{G}_{2008}^F$, $\mathcal{G}_{2008}^M$, $\mathcal{G}_{2013}^F$ and $\mathcal{G}_{2013}^M$ with a hierarchical random graph model, following \cite{clauset2008hierarchical} never give any statistical significance to an ordered structure in the network. The same result is obtained trimming the edge value function: erasing link with a weight below the median or the mean number of incoming or outgoing students immediately generates a network made of isolated sub-components.

The directed network $\mathcal{N}$ can also be explored in its weighted version, by accounting for the flow of students that each link has. By looking at the country level this time, the analysis allows to understand how, if on the one hand the number of participant countries has been stable in the last decade, as pictured in Figure \ref{fig:fig1}, the level of participation of each country is instead pretty heterogeneous, with student flows accounting for different proportion over the number of students enrolled at university. Table \ref{tab:tab1} shows the percentage indegree and outdegree measures for each country, weighted by the flow of incoming and outgoing students, averaged over the years 2008-2013 and normalized to the total number of students enrolled in higher education. For the incoming flows, the weighted indegree measure highlights the role of countries such as Spain, France, Great Britain, Germany and Italy as destination for European university students; while for the outgoing flows, the weighted outdegree measure draws attention to the role of small countries, such as Liechtenstein and Luxembourg, where a considerable part of the student population enrolled in higher education participates in the program.

\begin{table}[h!]
\centering
	\caption{Weighted average indegree and outdegree for the years from 2008 to 2013, normalized respectively to the average number of students enrolled in higher education in universities abroad and the amount of students enrolled at university in the country of reference.}
	\begin{tabular}{lcc}
		\hline
		& \multicolumn{1}{c}{Average weighted} & \multicolumn{1}{c}{Average weighted} \\
		& \multicolumn{1}{c}{indegree normalized} & \multicolumn{1}{c}{outdegree normalized} \\
		\hline
		Austria & 0.61  & 1.03 \\
		Belgium & 0.80  & 1.16 \\
		Bulgaria & 0.07  & 0.51 \\
		Croatia & 0.05  & 0.40 \\
		Cyprus & 0.05  & 0.71 \\
		Czech Republic & 0.64  & 1.35 \\
		Denmark & 0.74  & 0.73 \\
		Estonia & 0.11  & 1.11 \\
		Finland & 0.83  & 1.27 \\
		France & 3.31  & 1.09 \\
		Germany & 2.97  & 0.95 \\
		Great Britain & 2.51  & 0.37 \\
		Greece & 0.25  & 0.46 \\
		Hungary & 0.41  & 0.93 \\
		Iceland & 0.06  & 1.14 \\
		Ireland & 0.57  & 0.91 \\
		Italy & 2.34  & 1.07 \\
		Latvia & 0.09  & 1.40 \\
		Liechtenstein & 0.00  & 3.06 \\
		Lithuania & 0.20  & 1.53 \\
		Luxembourg & 0.01  & 6.52 \\
		Malta & 0.06  & 1.03 \\
		Norway & 0.49  & 0.57 \\
		Poland & 1.07  & 0.62 \\
		Portugal & 1.01  & 1.41 \\
		Republic of North Macedonia & - & 0.15 \\
		Romania & 0.18  & 0.54 \\
		Slovakia & 0.14  & 1.01 \\
		Slovenia & 0.18  & 1.27 \\
		Spain & 4.28  & 1.54 \\
		Sweden & 1.20  & 0.68 \\
		Switzerland & 0.26  & 0.93 \\
		The Netherlands & 1.05  & 0.90 \\
		Turkey & 0.67  & 0.20 \\
		\hline
	\end{tabular}%
	\label{tab:tab1}%
\end{table}%

\section*{Degree Distribution Analysis}
By considering the non-weighted Erasmus network of universities participating in the program, we analyse the degree distribution of this network by gender and along time, using 2008 and 2013 as initial and final benchmark years. Two universities are connected by a female-link (male-link) if there is at least one female (male) student moving from one university to the other one. 
As the network is directed, we observe both the indegree and the outdegree distribution. As common in most real networks, the degree distribution appears to be right skewed in all cases. Tables \ref{tab:tab2} and \ref{tab:tab3} collect summary statistics for every distribution considered together with the estimated values of skewness and kurtosis. The degree distributions appear to be far from the normal or the exponential model, as instead observed in \cite{derzsi2011topology} and thus they can be identified as ``heavy-tailed''\cite{clementi2016heavy}. 
\begin{table}[h!]
\centering
  \caption{Summary statistics and estimated skewness and kurtosis for the female indegree and outdegree distributions in 2008 and 2013.}
    \begin{tabular}{crccccccc}
\hline        & \multicolumn{1}{r}{} & \multicolumn{7}{c}{Female} \\
\hline        & \multicolumn{1}{r}{} & \multicolumn{1}{c}{min} & \multicolumn{1}{c}{median} & \multicolumn{1}{c}{mean} & \multicolumn{1}{c}{max} & \multicolumn{1}{c}{estimated} & \multicolumn{1}{c}{estimated} & \multicolumn{1}{c}{estimated} \\
& \multicolumn{1}{r}{} & \multicolumn{1}{c}{} & \multicolumn{1}{c}{} & \multicolumn{1}{c}{} & \multicolumn{1}{c}{} & \multicolumn{1}{c}{sd} & \multicolumn{1}{c}{skewness} & \multicolumn{1}{c}{kurtosis} \\
\hline
    \multirow{2}[1]{*}{Indegree} & 2008  & 0     & 2     & 18.86 & 669   & 46.26 & 5.81  & 53.03 \\
          & 2013  & 0     & 4     & 23.57 & 668   & 52.15 & 4.81  & 37.01 \\
    \multirow{2}[1]{*}{Outdegree} & 2008  & 0     & 3     & 18.86 & 537   & 41.91 & 4.59  & 33.70 \\
          & 2013  & 0     & 4     & 23.57 & 622   & 50.96 & 4.47  & 31.36 \\
\hline
    \end{tabular}%
  \label{tab:tab2}%
\end{table}%
\begin{table}[h!]
\centering
	\caption{Summary statistics and estimated skewness and kurtosis for the male indegree and outdegree distributions in 2008 and 2013.}
	\begin{tabular}{crccccccc}
		\hline        & \multicolumn{1}{r}{} & \multicolumn{7}{c}{Male} \\
		\hline        & \multicolumn{1}{r}{} & \multicolumn{1}{c}{min} & \multicolumn{1}{c}{median} & \multicolumn{1}{c}{mean} & \multicolumn{1}{c}{max} & \multicolumn{1}{c}{estimated} & \multicolumn{1}{c}{estimated} & \multicolumn{1}{c}{estimated} \\
		& \multicolumn{1}{r}{} & \multicolumn{1}{c}{} & \multicolumn{1}{c}{} & \multicolumn{1}{c}{} & \multicolumn{1}{c}{} & \multicolumn{1}{c}{sd} & \multicolumn{1}{c}{skewness} & \multicolumn{1}{c}{kurtosis} \\
		\hline
		\multirow{2}[1]{*}{Indegree} & 2008  & 0   & 2   & 13.63  & 403   & 32.09 & 4.96  & 37.79 \\
		& 2013  & 0     & 3     & 16.98 & 392   & 36.19 & 4.19  & 27.06 \\
		\multirow{2}[1]{*}{Outdegree} & 2008  & 0     & 2     & 13.63 & 541   & 31.89  & 5.63  & 54.78 \\
		& 2013  & 0     & 3     & 16.98 & 484   & 36.91 & 4.63  & 34.13 \\
		\hline
	\end{tabular}%
	\label{tab:tab3}%
\end{table}%

\FloatBarrier
\subsection*{Comparison along time}
\begin{figure}[h!]	
\centering
\begin{subfigure}[b]{0.45\linewidth}
\centering
	\includegraphics[scale=0.25]{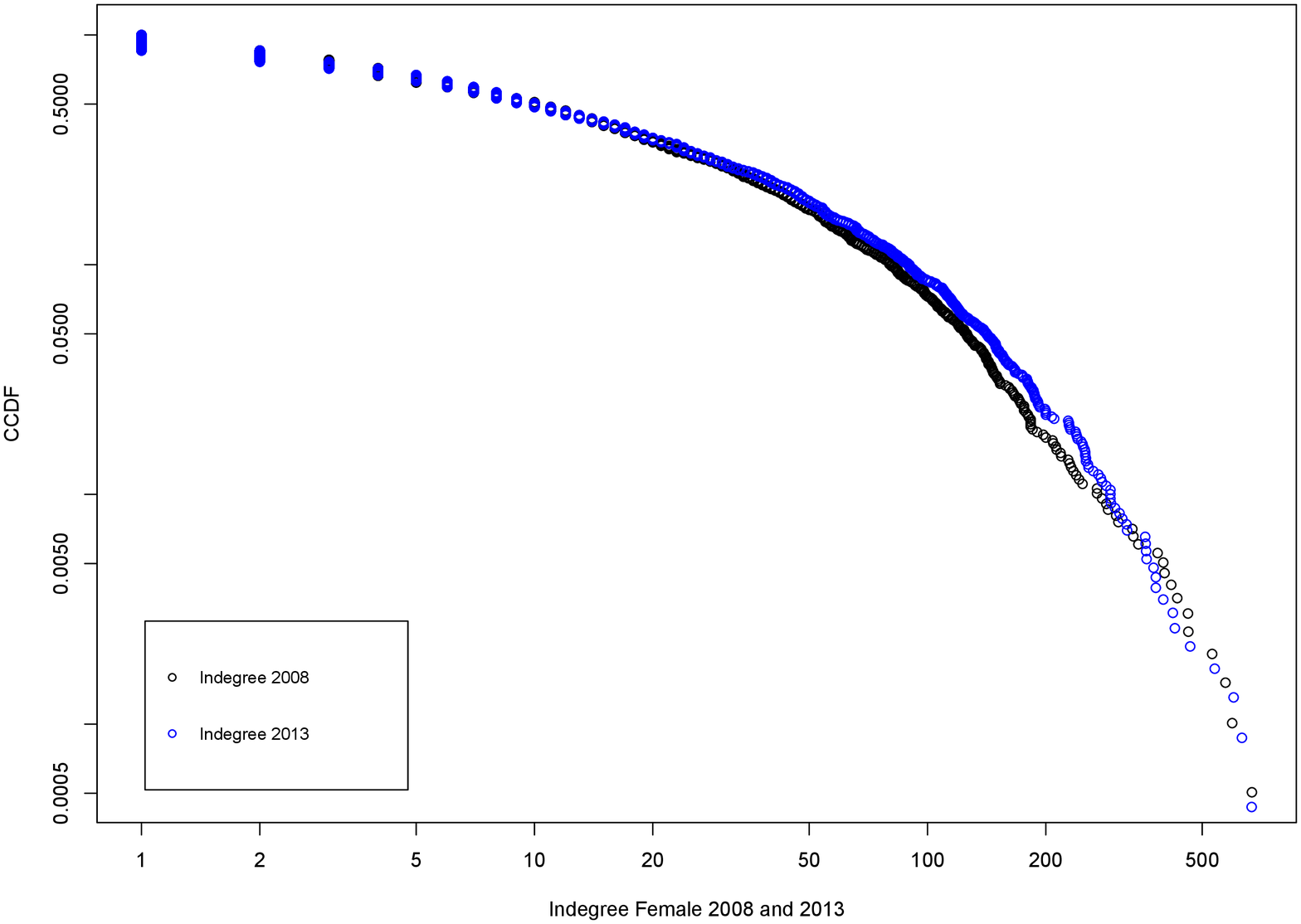}
	\caption{}
	\end{subfigure}
\begin{subfigure}[b]{0.45\linewidth}
\centering
	\includegraphics[scale=0.25]{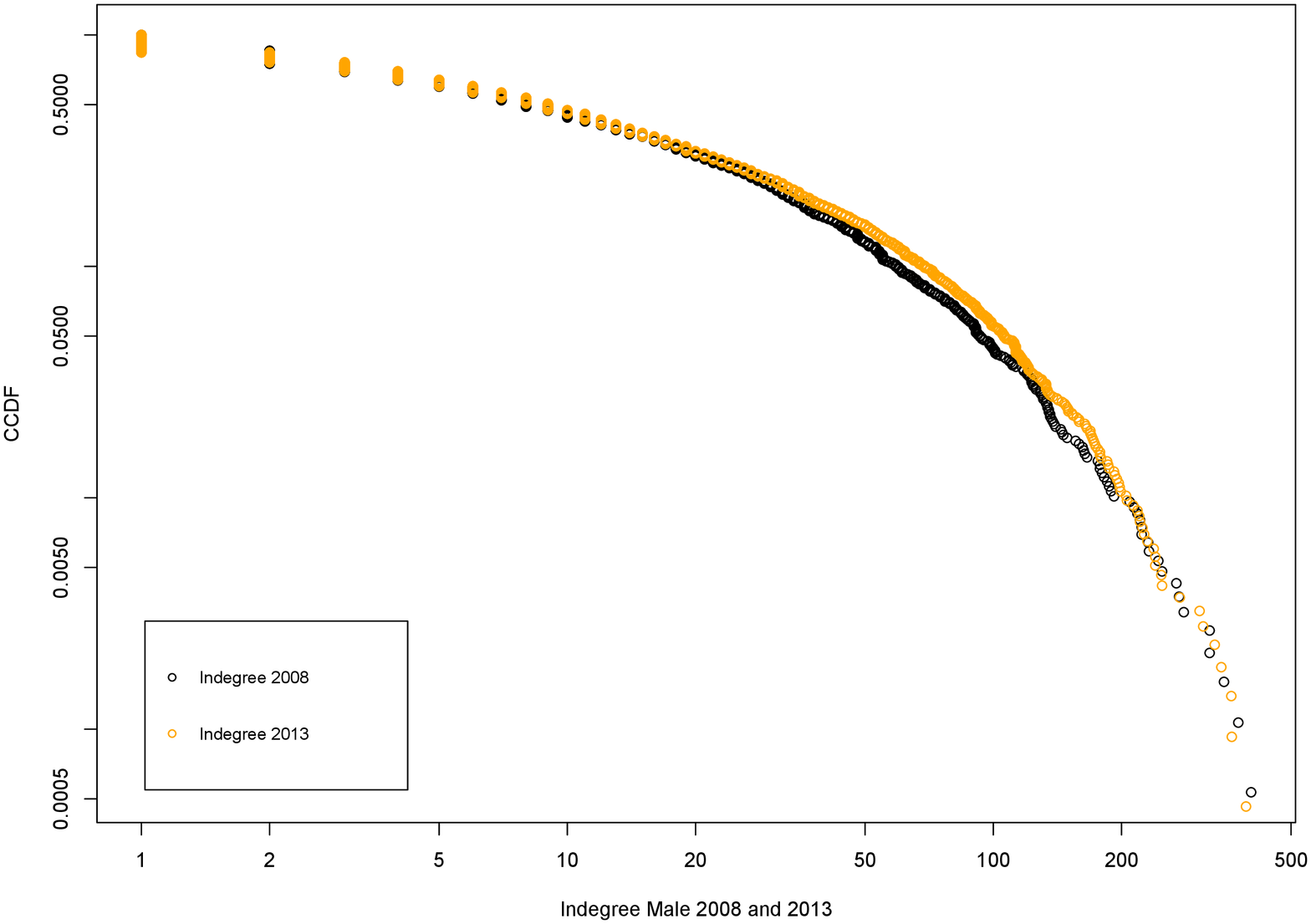}
	\caption{}
	\end{subfigure}
	\caption{Female and male indegree distribution compared by year.
		Complementary cumulative distribution function (CCDF) on log-log scale of the female (a) and male (b) indegree distribution for the years 2008 and 2013}
			\label{fig:fig5}
\end{figure}

\begin{figure}[h!]
\centering
\begin{subfigure}[b]{0.45\linewidth}
\centering
	\includegraphics[scale=0.25]{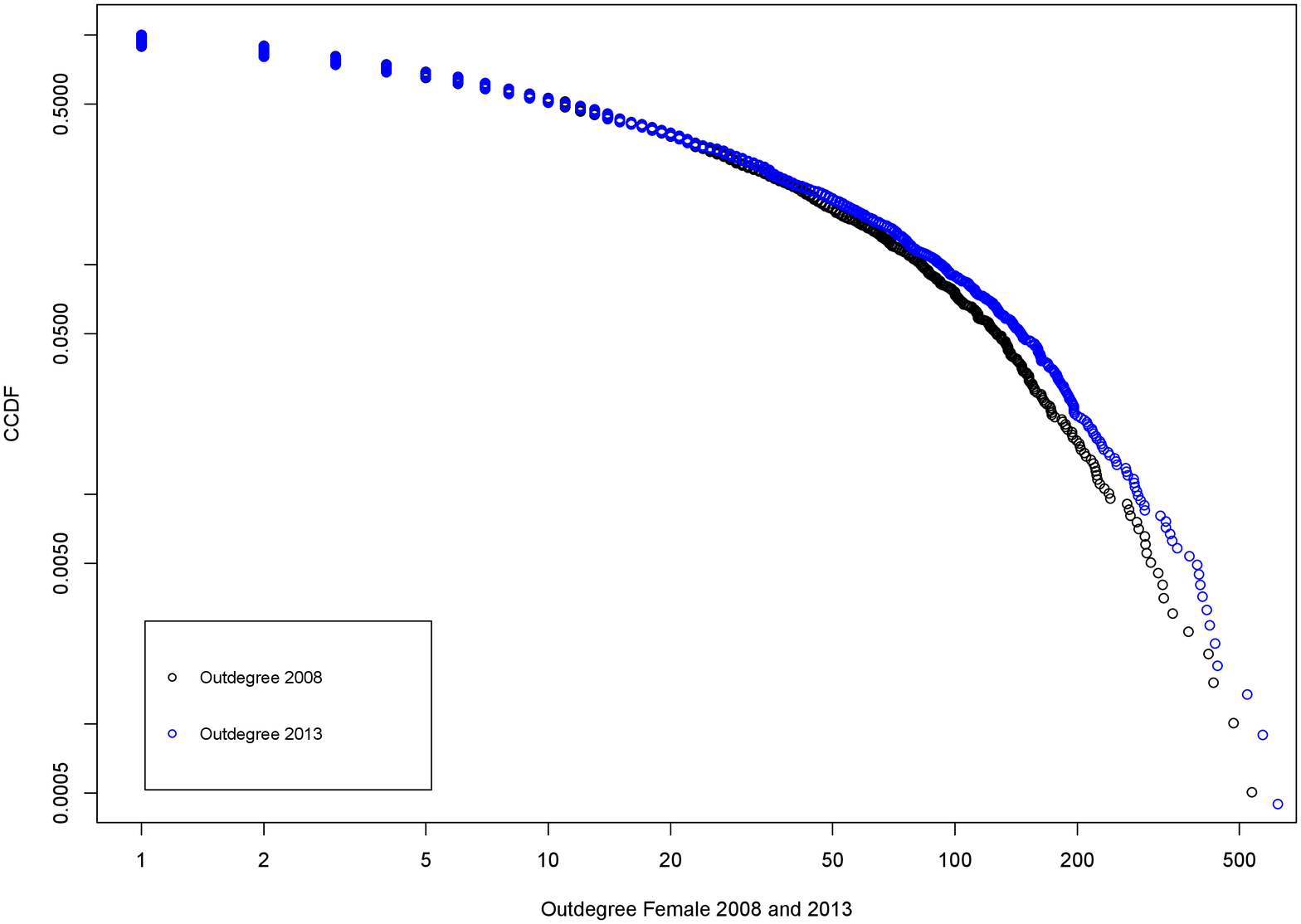}
	\caption{}
	\end{subfigure}
\begin{subfigure}[b]{0.45\linewidth}
\centering
	\includegraphics[scale=0.25]{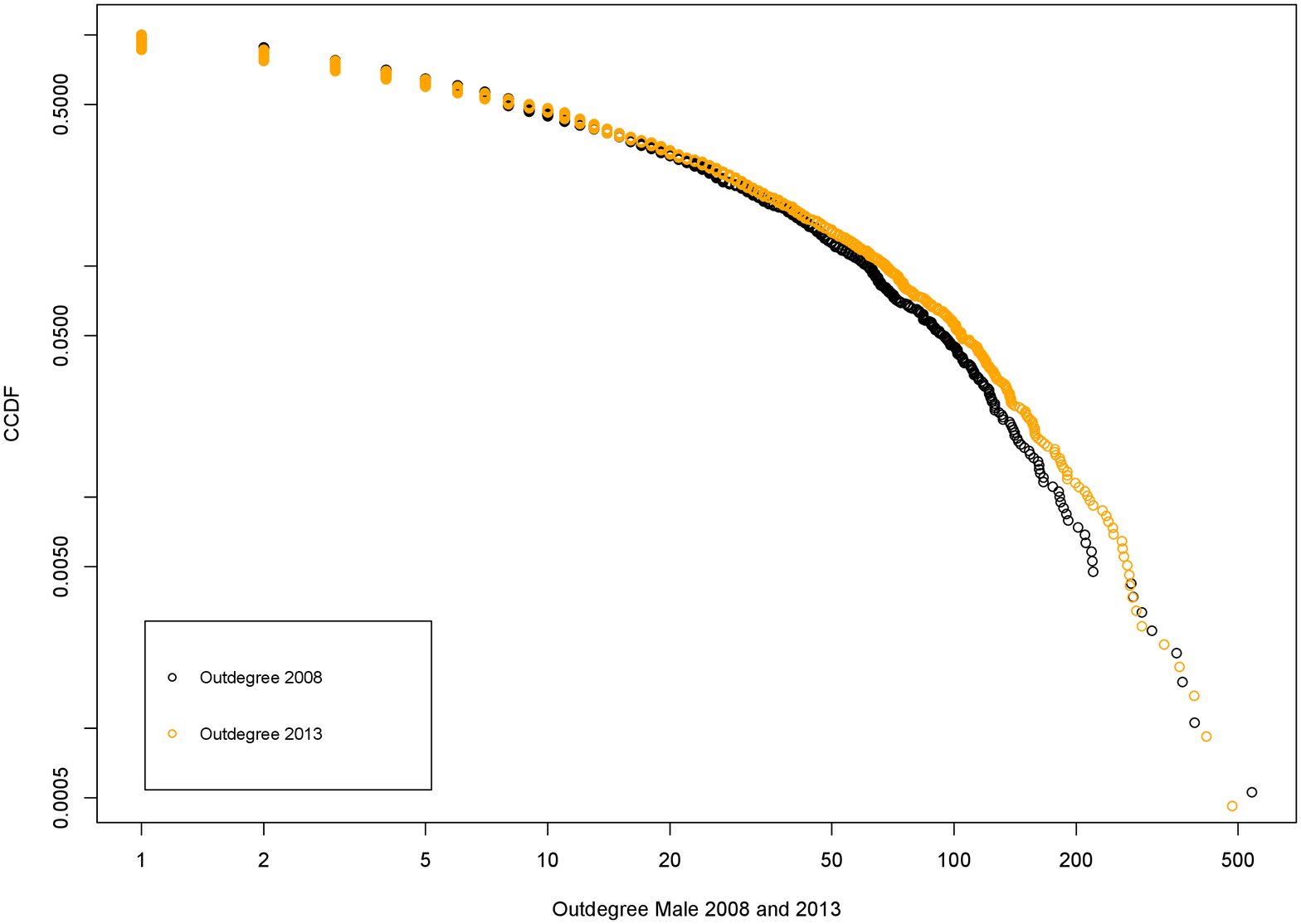}
	\caption{}
	\end{subfigure}
	\caption{Female and male outdegree distribution compared by year.
		Complementary cumulative distribution function (CCDF) on log-log scale of the female (a) and male (b) outdegree distribution for the years 2008 and 2013}
	\label{fig:fig6}
\end{figure}

Figure \ref{fig:fig5} and Figure \ref{fig:fig6} compare the complementary cumulative distribution function (CCDF), $Pr(X)\geq x$, on a log-log scale respectively of the indegree and the outdegree for the years 2008 and 2013, for male and female links. Looking at the indegree distributions, both 2008 data are positioned under the 2013 and, especially for the female case, the CCDF corresponding to 2013 appears "fatter" in its central part, while the former looks more stretched. The tail of the distribution appears to have squeezed along the years considered. 
This is not visible for the outdegree distributions, which seem to have maintained the same shape along time.

\FloatBarrier
\subsection*{Comparison between genders}
\begin{figure}[h!]
\centering
\begin{subfigure}[b]{0.45\linewidth}
\centering
	\includegraphics[scale=0.25]{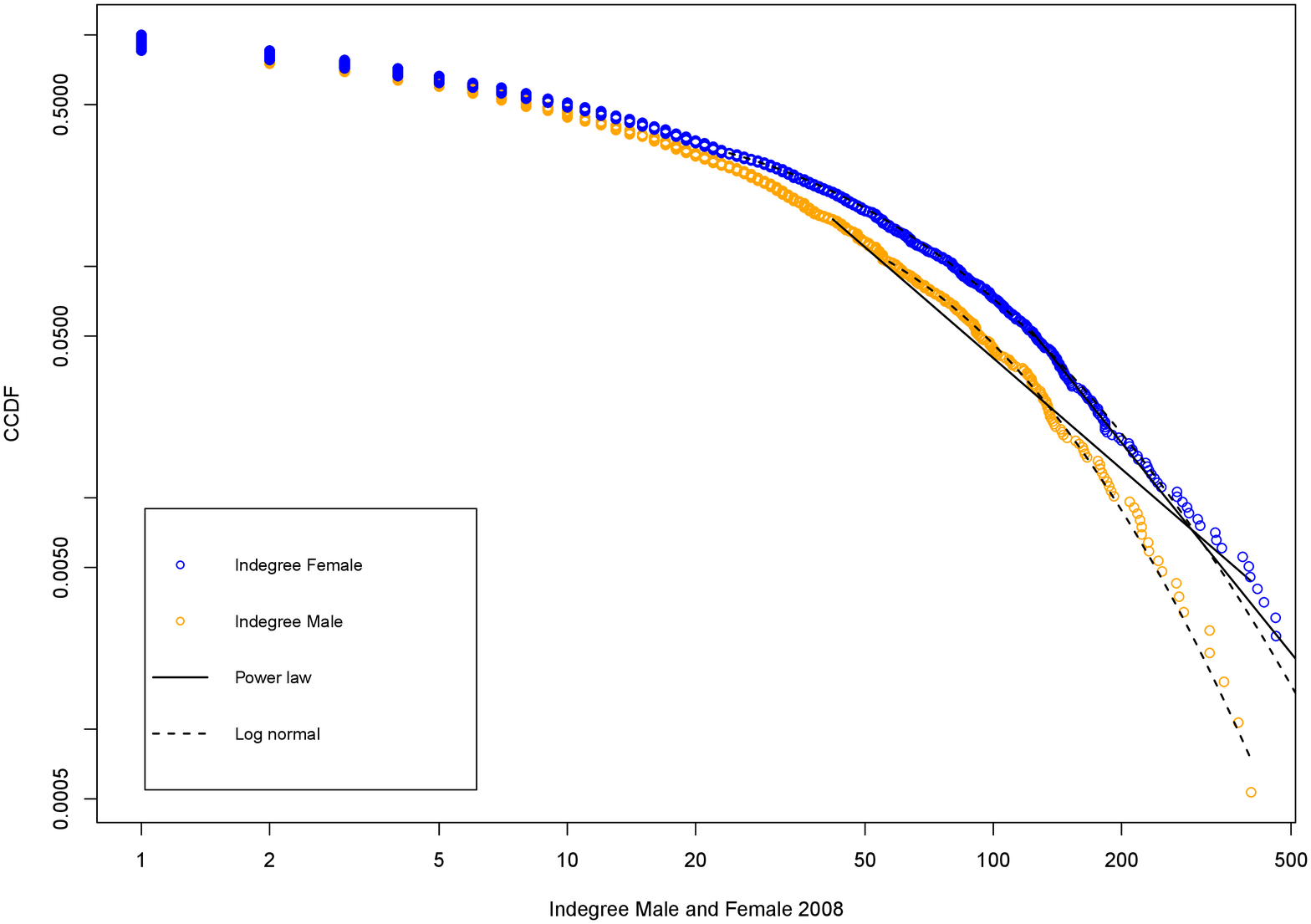}
	\caption{}
	\end{subfigure}
\begin{subfigure}[b]{0.45\linewidth}
\centering
	\includegraphics[scale=0.25]{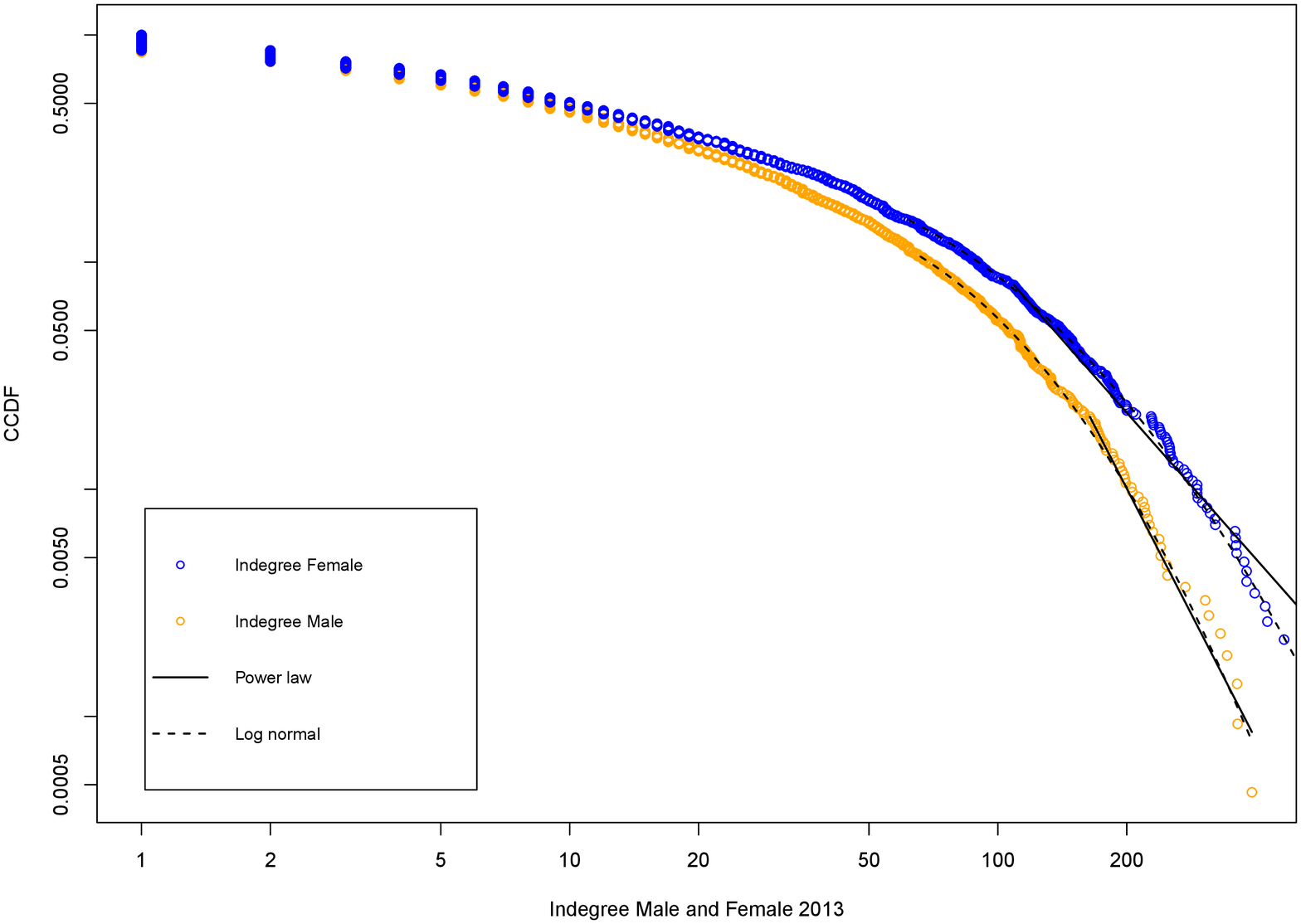}
	\caption{}
	\end{subfigure}
	\caption{2008 and 2013 indegree distributions compared by gender.
		Complementary cumulative distribution function (CCDF) on log-log scale of the female and male indegree distribution for the years 2008 (a) and 2013 (b). For each empirical distribution, the plots display also the lines of fit for a power law and a log normal distributional model.}
	\label{fig:fig7}
\end{figure}

\begin{figure}[h!]
\centering
\begin{subfigure}[b]{0.45\linewidth}
\centering
	\includegraphics[scale=0.25]{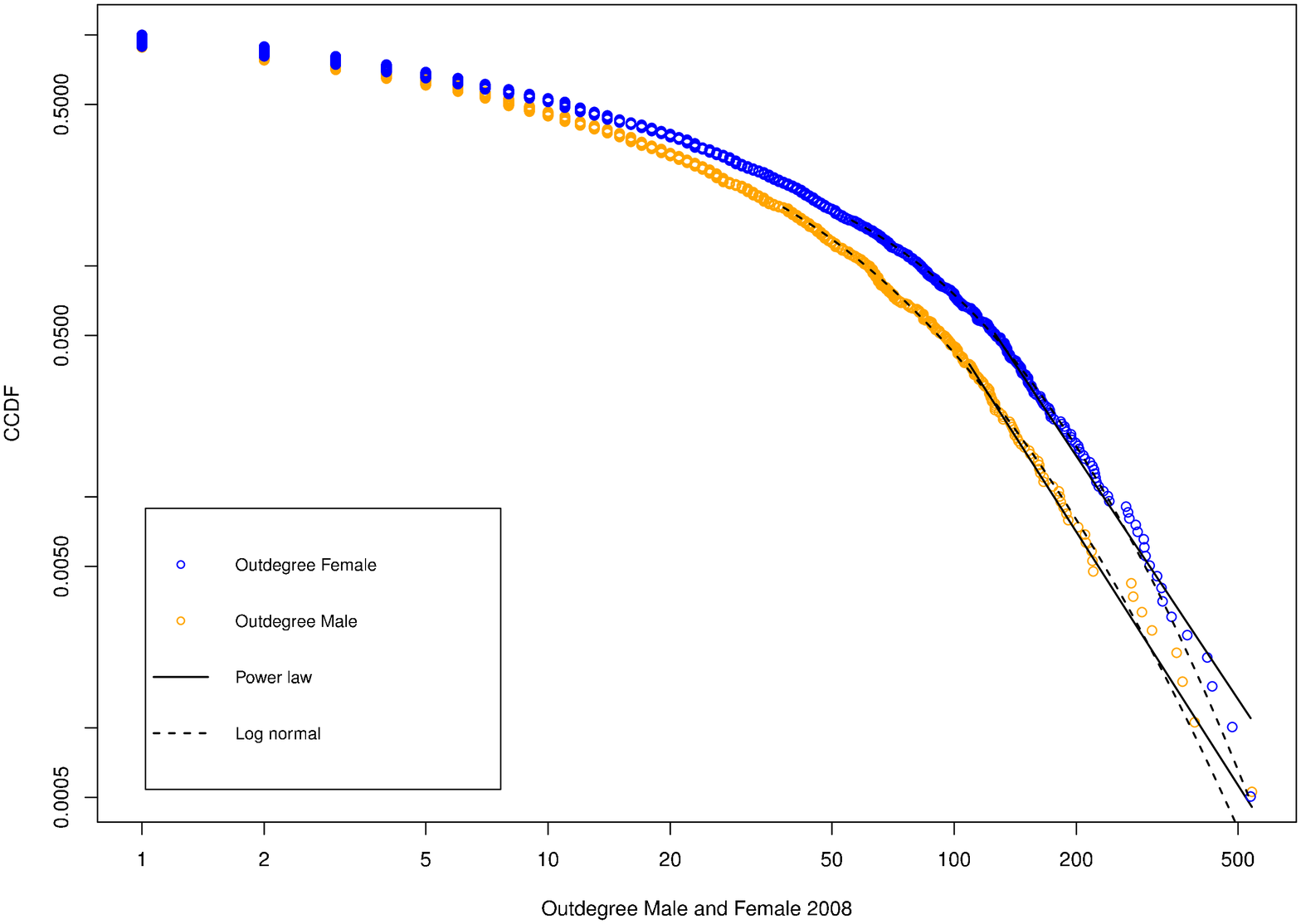}
	\caption{}
	\end{subfigure}
\begin{subfigure}[b]{0.45\linewidth}
\centering
	\includegraphics[scale=0.25]{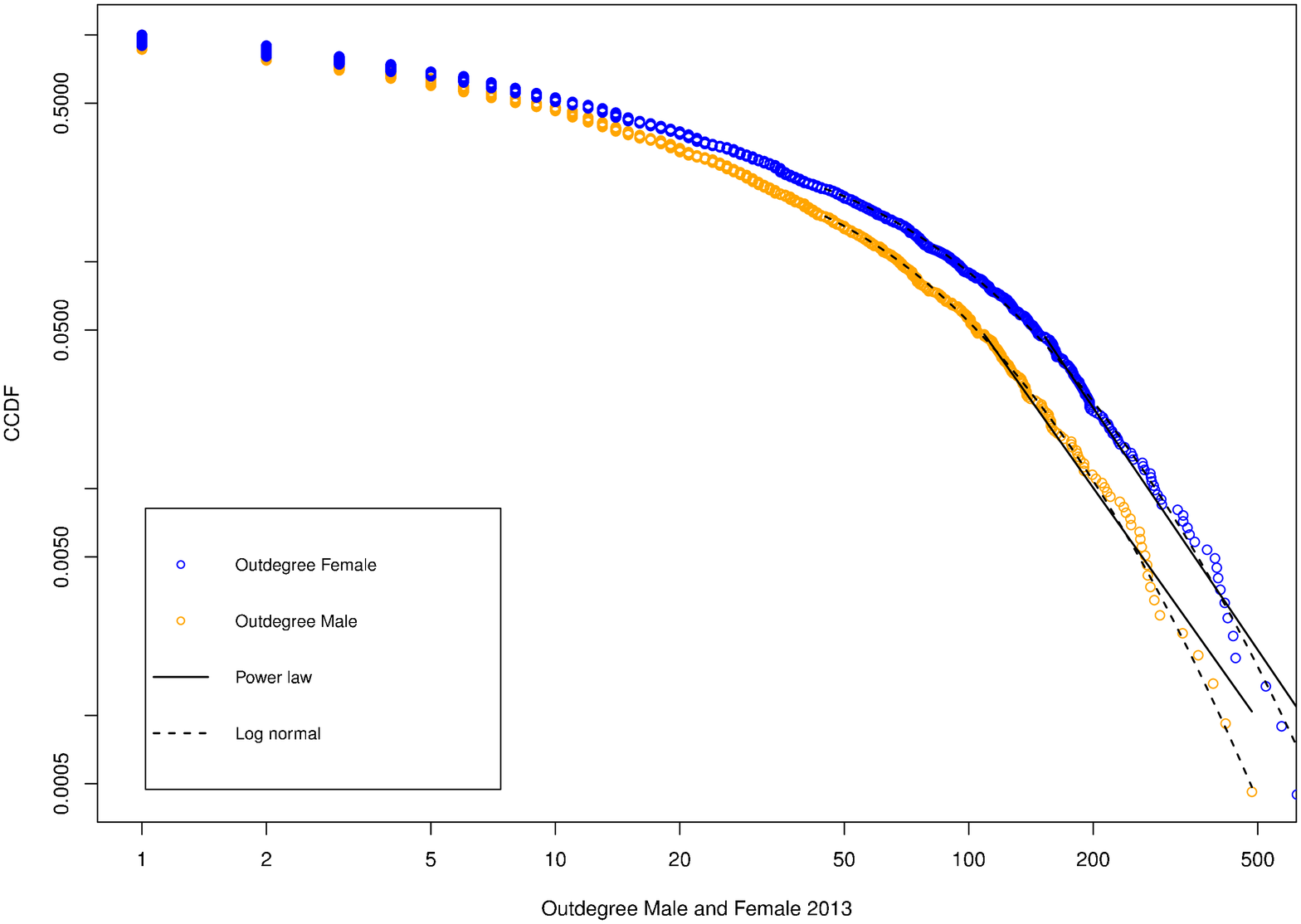}
	\caption{}
	\end{subfigure}
	\caption{2008 and 2013 outdegree distributions compared by gender
		Complementary cumulative distribution function (CCDF) on log-log scale of the female and male outdegree distribution for the years 2008 (a) and 2013 (b). For each empirical distribution, the plots display also the lines of fit for a power law and a log normal distributional model.}
	\label{fig:fig8}
\end{figure}

The same graphs are reported in Figures \ref{fig:fig7} and \ref{fig:fig8}, this time comparing the distribution for male and female connections in the same plot. The plotted densities include now also lines of fit for a power law distributional model and a log normal distribution.
A power law degree distribution, $p(x)\propto x^{-\alpha}$, is observed in the so-called scale-free networks \cite{barabasi1999mean}, although the empirical distribution usually follows a power law model only in its upper tail, i.e. starting from a threshold $x_{min}$. 

The routine employed for fitting heavy-tailed distributions is developed by \cite{gillespie2015fitting} and based on \cite{clauset2009power}; it relies on maximum likelihood estimators and goodness-of-fit based approach to determine the cut-off $x_{min}$. In particular, the optimal choice of $x_{min}$ is determined by minimizing the distance $D$ between the probability distribution of the data and the best-fit power law model, measured by the Kolmogorov-Smirnov (K-S) statistic:

$$
D = \max_{x\geq x_{min}} \mid S(x)-P(x)\mid,
$$

where $S(x)$ is the Cumulative Distribution Function (CDF) of the data and $P(x)$ is the CDF for the power-law fitted model.

Following the framework proposed by \cite{clauset2009power}, the visual inspection and the distribution fit are complemented with a goodness-of-fit test, based again on the K-S. 
Via bootstrapping, a distribution of the K-S statistic is generated following the creation of a large number of power-law distributed synthetic data sets with scaling parameter $\alpha$ and lower bound $x_{min}$ equal to those of the distribution that best fits the observed data. A p-value is thus generated as the fraction of the time that the K-S statistic is larger than its value for the empirical data. The p-value quantifies the following hypotheses:
$$
\begin{array}{l}
H_0: \text{the power law fitted model is a plausible option} \\
H_{1}: \text{the power law fitted model is not a plausible option}
\end{array}
$$

Table \ref{tab:tab4} collects the estimated values for the scaling parameter $\alpha$ and the threshold value $x_{min}$ for each power law fitted model and shows the relative K-S statistic together with the \textit{p-value} resulting from the goodness-of-fit test for each case.
As in \cite{clauset2009power} we decide to rule out the power law hypotesis if $p \leq0.1$; this is the case for the female indegree distribution in 2013 and the male indegree distribution in 2008, which therefore are not well described by a power law. 

\begin{table}[h!]
  \centering
  \caption{Estimated parameter and lower bound for a power model fitted model with relative K-S statistic and p-value for the goodness-of-fit test.}
    \begin{tabular}{rlcccc}
    \hline
          &       & $x_{min}$  & $\alpha$     & K-S   & p-value \\
    \hline
    \multicolumn{1}{l}{Female} & Indegree 2008 & 123   & 3.28  & 0.035 & 0.87 \\
          & Indegree2013 & 109   & 3.13  & 0.064 & 0.02 \\
          & Outdegree 2008 & 126   & 3.64  & 0.046 & 0.68 \\
          & Outdegree 2013 & 153   & 3.68  & 0.045 & 0.66 \\
    \multicolumn{1}{l}{Male} & Indegree 2008 & 42    & 2.59  & 0.073 & 0 \\
          & Indegree 2013 & 164   & 4.66  & 0.070 & 0.32 \\
          & Outdegree 2008 & 109   & 3.75  & 0.041 & 0.97 \\
          & Outdegree 2013 & 109   & 3.56  & 0.049 & 0.48 \\
    \hline
    \end{tabular}%
  \label{tab:tab4}%
\end{table}%

For the sake of accuracy, the analysis is complemented with a Vuong's test comparing the power law fit with a log normal distribution fit, which suggests that there is not a preferred model between those tested \cite{vuong1989likelihood}. 

Although it is not trivial to identify the best distributional model for the degree distributions, some observations can be drawn. For the indegree distribution we observe an opposite behaviour along time in the tail of the female and male distribution. The former has squeezed along the years considered, so that the test conducted supports the hypothesis that the power law model is no longer a good description for the tail of this distribution. On the contrary, the male indegree distribution shows a ``heavy'' but not ``fat'' tail in 2008, meaning that it goes to zero slower than an exponential model but faster than a power law; in 2013, instead, the tail of the male indegree distribution seems to have stretched and proves to be well described by a power law. This change in the distributional model suggests that starting from a situation characterized by strong gender imbalance in favor of female connections, the Erasmus network has been moving towards an increased gender parity in its incoming connections.

On the other hand, the outdegree distributions remain pretty stable over time and do not behave much differently by gender. We can hypothesize that when new universities adhere to the program, they follow a mechanism of preferential attachment, as described in \cite{derzsi2011topology}: they connect with the most popular hubs, with high indegree values, not necessarily receiving a connection in the opposite direction, so that the outdegree value increases for those observation located in the core of the distribution while not affecting the behaviour of the tail. 

In general, the results of our analysis suggest that the degree distribution of the Erasmus network has changed along time and findings by \cite{derzsi2011topology} do not hold true with the most recent data. 

\section*{Conclusions}
The Erasmus Program is characterized by a strong gender bias in favour of female students. This work quantifies the gender bias in the Erasmus program between 2008 and 2013, using novel data at the university level from the EU open data portal. After describing the structure of the program in great details, carrying out a descriptive analysis across fields of study, and identifying key universities as senders and receivers, the paper gives emphasis to the multi-dimensionality of the gender bias across countries and fields of study. It is evident that with minimal oscillations along the years, the gender bias persisted over time, with a proportion of female over men of 1.38 to 1. This is due to the denser network of connections involving female students, that prevail in fields such as Arts and Humanities, and Business Administration and Law. The paper, also, tests the difference in the degree distribution of the Erasmus network along time and between genders, giving evidence of a higher prevalence of universities in the female Erasmus network receiving higher level of inflows respect to the one of the male Erasmus network. Finally, some evidence of change is shown along time: the bias in favor of female students is strongly reduced especially in fields such as Social Sciences, journalism and information; Information and communication technologies; Health and welfare; and in Business Administration and Law. These changes are more relevant in Eastern European and Mediterranean countries and can foster the convergence of male and female students flows possibly resulting in a future reduction of the bias after thirty years.  
\section*{Competing interests}
  The authors declare that they have no competing interests.

\section*{Author's contributions}
  Both authors planned the research project, Silvia Leoni collected the original data, Luca De Benedictis prepared the data sets and coded the network analysis in R, Silvia Leoni ran the degree distribution analysis, both authors contributed to the empirical analysis and to the writing of the paper. All data, R script and Stata do files are available for replication purposes.

\section*{Acknowledgements}
  The authors wish to thank Domenico Vistocco for the kind tutorial on how to clean and wrangle the Erasmus data.

\bibliographystyle{bmc-mathphys} 
\bibliography{Gender_bias}


\begin{thebibliography}{21}
\ifx \bisbn   \undefined \def \bisbn  #1{ISBN #1}\fi
\ifx \binits  \undefined \def \binits#1{#1}\fi
\ifx \bauthor  \undefined \def \bauthor#1{#1}\fi
\ifx \batitle  \undefined \def \batitle#1{#1}\fi
\ifx \bjtitle  \undefined \def \bjtitle#1{#1}\fi
\ifx \bvolume  \undefined \def \bvolume#1{\textbf{#1}}\fi
\ifx \byear  \undefined \def \byear#1{#1}\fi
\ifx \bissue  \undefined \def \bissue#1{#1}\fi
\ifx \bfpage  \undefined \def \bfpage#1{#1}\fi
\ifx \blpage  \undefined \def \blpage #1{#1}\fi
\ifx \burl  \undefined \def \burl#1{\textsf{#1}}\fi
\ifx \doiurl  \undefined \def \doiurl#1{\textsf{#1}}\fi
\ifx \betal  \undefined \def \betal{\textit{et al.}}\fi
\ifx \binstitute  \undefined \def \binstitute#1{#1}\fi
\ifx \binstitutionaled  \undefined \def \binstitutionaled#1{#1}\fi
\ifx \bctitle  \undefined \def \bctitle#1{#1}\fi
\ifx \beditor  \undefined \def \beditor#1{#1}\fi
\ifx \bpublisher  \undefined \def \bpublisher#1{#1}\fi
\ifx \bbtitle  \undefined \def \bbtitle#1{#1}\fi
\ifx \bedition  \undefined \def \bedition#1{#1}\fi
\ifx \bseriesno  \undefined \def \bseriesno#1{#1}\fi
\ifx \blocation  \undefined \def \blocation#1{#1}\fi
\ifx \bsertitle  \undefined \def \bsertitle#1{#1}\fi
\ifx \bsnm \undefined \def \bsnm#1{#1}\fi
\ifx \bsuffix \undefined \def \bsuffix#1{#1}\fi
\ifx \bparticle \undefined \def \bparticle#1{#1}\fi
\ifx \barticle \undefined \def \barticle#1{#1}\fi
\ifx \bconfdate \undefined \def \bconfdate #1{#1}\fi
\ifx \botherref \undefined \def \botherref #1{#1}\fi
\ifx \url \undefined \def \url#1{\textsf{#1}}\fi
\ifx \bchapter \undefined \def \bchapter#1{#1}\fi
\ifx \bbook \undefined \def \bbook#1{#1}\fi
\ifx \bcomment \undefined \def \bcomment#1{#1}\fi
\ifx \oauthor \undefined \def \oauthor#1{#1}\fi
\ifx \citeauthoryear \undefined \def \citeauthoryear#1{#1}\fi
\ifx \endbibitem  \undefined \def \endbibitem {}\fi
\ifx \bconflocation  \undefined \def \bconflocation#1{#1}\fi
\ifx \arxivurl  \undefined \def \arxivurl#1{\textsf{#1}}\fi
\csname PreBibitemsHook\endcsname

\bibitem{maiworm2001erasmus}
\begin{barticle}
\bauthor{\bsnm{Maiworm}, \binits{F.}}:
\batitle{Erasmus: continuity and change in the 1990s}.
\bjtitle{European journal of education}
\bvolume{36}(\bissue{4}),
\bfpage{459}--\blpage{472}
(\byear{2001})
\end{barticle}
\endbibitem

\bibitem{bottcher2016gender}
\begin{botherref}
\oauthor{\bsnm{Bottcher}, \binits{L.}},
\oauthor{\bsnm{Araujo}, \binits{N.A.}},
\oauthor{\bsnm{Nagler}, \binits{J.}},
\oauthor{\bsnm{Mendes}, \binits{J.F.}},
\oauthor{\bsnm{Helbing}, \binits{D.}},
\oauthor{\bsnm{Herrmann}, \binits{H.J.}}:
Gender gap in the erasmus mobility program.
PLoS ONE
\textbf{11}(2)
(2016)
\end{botherref}
\endbibitem

\bibitem{bhandari2017women}
\begin{botherref}
\oauthor{\bsnm{Bhandari}, \binits{R.}}:
Women on the move: Gender dimensions of academic mobility.
Institute of International Education
(2017)
\end{botherref}
\endbibitem

\bibitem{martin2014gender}
\begin{barticle}
\bauthor{\bsnm{Martin}, \binits{F.}}:
\batitle{The gender of mobility}.
\bjtitle{Intersections: Gender and Sexuality in Asia and the Pacific}
\bvolume{35},
\bfpage{33}--\blpage{47}
(\byear{2014})
\end{barticle}
\endbibitem

\bibitem{myers2019geography}
\begin{barticle}
\bauthor{\bsnm{Myers}, \binits{R.M.}},
\bauthor{\bsnm{Griffin}, \binits{A.L.}}:
\batitle{The geography of gender inequality in international higher education}.
\bjtitle{Journal of Studies in International Education}
\bvolume{23}(\bissue{4}),
\bfpage{429}--\blpage{450}
(\byear{2019})
\end{barticle}
\endbibitem

\bibitem{faggian2007some}
\begin{barticle}
\bauthor{\bsnm{Faggian}, \binits{A.}},
\bauthor{\bsnm{McCann}, \binits{P.}},
\bauthor{\bsnm{Sheppard}, \binits{S.C.}}:
\batitle{Some evidence that women are more mobile than men: Gender differences
  in uk graduate migration behavior}.
\bjtitle{Journal of Regional Science}
\bvolume{47}(\bissue{3}),
\bfpage{517}--\blpage{539}
(\byear{2007})
\end{barticle}
\endbibitem

\bibitem{derzsi2011topology}
\begin{barticle}
\bauthor{\bsnm{Derzsi}, \binits{A.}},
\bauthor{\bsnm{Derzsy}, \binits{N.}},
\bauthor{\bsnm{K{\'a}ptalan}, \binits{E.}},
\bauthor{\bsnm{N{\'e}da}, \binits{Z.}}:
\batitle{Topology of the erasmus student mobility network}.
\bjtitle{Physica A: Statistical Mechanics and its Applications}
\bvolume{390}(\bissue{13}),
\bfpage{2601}--\blpage{2610}
(\byear{2011})
\end{barticle}
\endbibitem

\bibitem{uis2012international}
\begin{botherref}
\oauthor{\bsnm{UIS}}:
International standard classification of education: Isced 2011.
Technical report,
UNESCO Institute for Statistics, Montreal
(2012)
\end{botherref}
\endbibitem

\bibitem{unesco2014isced}
\begin{botherref}
\oauthor{\bsnm{UIS}}:
Isced fields of education and training 2013 (isced-f 2013): manual to accompany
  the international standard classification of education.
Technical report,
UNESCO Institute for Statistics, Montreal, Quebec
(2014)
\end{botherref}
\endbibitem

\bibitem{corradi2015}
\begin{bbook}
\bauthor{\bsnm{Corradi}, \binits{S.}}:
\bbtitle{Student Mobility in Higher Education. Erasmus and Erasmus Plus}.
\bpublisher{Laboratory of Lifelong Learning Department of Education and
  Training “Roma Tre” State University},
\blocation{Rome}
(\byear{2015})
\end{bbook}
\endbibitem

\bibitem{european2015erasmus+}
\begin{botherref}
\oauthor{\bsnm{Commission}, \binits{E.}}:
Erasmus+ Programme: Annual Report 2014.
Publications Office of the European Union Luxembourg
(2015)
\end{botherref}
\endbibitem

\bibitem{europeancommission2018}
\begin{botherref}
\oauthor{\bsnm{Commission}, \binits{E.}}:
Investing in people. Making Erasmus even better.
Publications Office of the European Union Luxembourg
(2018)
\end{botherref}
\endbibitem

\bibitem{benedictis2005three}
\begin{barticle}
\bauthor{\bsnm{De~Benedictis}, \binits{L.}}:
\batitle{Three decades of italian comparative advantages}.
\bjtitle{World Economy}
\bvolume{28}(\bissue{11}),
\bfpage{1679}--\blpage{1709}
(\byear{2005})
\end{barticle}
\endbibitem

\bibitem{botella2019gender}
\begin{barticle}
\bauthor{\bsnm{Botella}, \binits{C.}},
\bauthor{\bsnm{Rueda}, \binits{S.}},
\bauthor{\bsnm{L{\'o}pez-I{\~n}esta}, \binits{E.}},
\bauthor{\bsnm{Marzal}, \binits{P.}}:
\batitle{Gender diversity in stem disciplines: A multiple factor problem}.
\bjtitle{Entropy}
\bvolume{21}(\bissue{1}),
\bfpage{30}
(\byear{2019})
\end{barticle}
\endbibitem

\bibitem{oecd2017pursuit}
\begin{botherref}
\oauthor{\bsnm{OECD}}:
The pursuit of gender equality-an uphill battle.
Technical report,
OECD publishing
(2017)
\end{botherref}
\endbibitem

\bibitem{clauset2008hierarchical}
\begin{barticle}
\bauthor{\bsnm{Clauset}, \binits{A.}},
\bauthor{\bsnm{Moore}, \binits{C.}},
\bauthor{\bsnm{Newman}, \binits{M.E.}}:
\batitle{Hierarchical structure and the prediction of missing links in
  networks}.
\bjtitle{Nature}
\bvolume{453}(\bissue{7191}),
\bfpage{98}--\blpage{101}
(\byear{2008})
\end{barticle}
\endbibitem

\bibitem{clementi2016heavy}
\begin{bchapter}
\bauthor{\bsnm{Clementi}, \binits{F.}}:
\bctitle{Heavy-tailed distributions for agent-based economic modelling}.
In: \bbtitle{Economics with Heterogeneous Interacting Agents},
pp. \bfpage{157}--\blpage{190}.
\bpublisher{Springer},
\blocation{Switzerland}
(\byear{2016})
\end{bchapter}
\endbibitem

\bibitem{barabasi1999mean}
\begin{barticle}
\bauthor{\bsnm{Barab{\'a}si}, \binits{A.-L.}},
\bauthor{\bsnm{Albert}, \binits{R.}},
\bauthor{\bsnm{Jeong}, \binits{H.}}:
\batitle{Mean-field theory for scale-free random networks}.
\bjtitle{Physica A: Statistical Mechanics and its Applications}
\bvolume{272}(\bissue{1-2}),
\bfpage{173}--\blpage{187}
(\byear{1999})
\end{barticle}
\endbibitem

\bibitem{gillespie2015fitting}
\begin{botherref}
\oauthor{\bsnm{Gillespie}, \binits{C.S.}}, et al.:
Fitting heavy tailed distributions: The powerlaw package.
Journal of Statistical Software
\textbf{64}(i02)
(2015)
\end{botherref}
\endbibitem

\bibitem{clauset2009power}
\begin{barticle}
\bauthor{\bsnm{Clauset}, \binits{A.}},
\bauthor{\bsnm{Shalizi}, \binits{C.R.}},
\bauthor{\bsnm{Newman}, \binits{M.E.}}:
\batitle{Power-law distributions in empirical data}.
\bjtitle{SIAM review}
\bvolume{51}(\bissue{4}),
\bfpage{661}--\blpage{703}
(\byear{2009})
\end{barticle}
\endbibitem

\bibitem{vuong1989likelihood}
\begin{botherref}
\oauthor{\bsnm{Vuong}, \binits{Q.H.}}:
Likelihood ratio tests for model selection and non-nested hypotheses.
Econometrica: Journal of the Econometric Society,
307--333
(1989)
\end{botherref}
\endbibitem

\end{thebibliography}

\newcommand{\BMCxmlcomment}[1]{}

\BMCxmlcomment{

<refgrp>

<bibl id="B1">
  <title><p>ERASMUS: continuity and change in the 1990s</p></title>
  <aug>
    <au><snm>Maiworm</snm><fnm>F</fnm></au>
  </aug>
  <source>European journal of education</source>
  <publisher>JSTOR</publisher>
  <pubdate>2001</pubdate>
  <volume>36</volume>
  <issue>4</issue>
  <fpage>459</fpage>
  <lpage>-472</lpage>
</bibl>

<bibl id="B2">
  <title><p>Gender Gap in the ERASMUS Mobility Program</p></title>
  <aug>
    <au><snm>Bottcher</snm><fnm>L</fnm></au>
    <au><snm>Araujo</snm><fnm>NA</fnm></au>
    <au><snm>Nagler</snm><fnm>J</fnm></au>
    <au><snm>Mendes</snm><fnm>JF</fnm></au>
    <au><snm>Helbing</snm><fnm>D</fnm></au>
    <au><snm>Herrmann</snm><fnm>HJ</fnm></au>
  </aug>
  <source>PLoS ONE</source>
  <publisher>Public Library of Science</publisher>
  <pubdate>2016</pubdate>
  <volume>11</volume>
  <issue>2</issue>
</bibl>

<bibl id="B3">
  <title><p>Women on the move: Gender Dimensions of Academic
  Mobility</p></title>
  <aug>
    <au><snm>Bhandari</snm><fnm>R</fnm></au>
  </aug>
  <source>Institute of International Education</source>
  <pubdate>2017</pubdate>
</bibl>

<bibl id="B4">
  <title><p>The Gender of Mobility</p></title>
  <aug>
    <au><snm>Martin</snm><fnm>F</fnm></au>
  </aug>
  <source>Intersections: Gender and Sexuality in Asia and the Pacific</source>
  <pubdate>2014</pubdate>
  <volume>35</volume>
  <fpage>33</fpage>
  <lpage>-47</lpage>
</bibl>

<bibl id="B5">
  <title><p>The Geography of Gender Inequality in International Higher
  Education</p></title>
  <aug>
    <au><snm>Myers</snm><fnm>RM</fnm></au>
    <au><snm>Griffin</snm><fnm>AL</fnm></au>
  </aug>
  <source>Journal of Studies in International Education</source>
  <publisher>SAGE Publications Sage CA: Los Angeles, CA</publisher>
  <pubdate>2019</pubdate>
  <volume>23</volume>
  <issue>4</issue>
  <fpage>429</fpage>
  <lpage>-450</lpage>
</bibl>

<bibl id="B6">
  <title><p>Some Evidence that Women are More Mobile than Men: Gender
  Differences in UK Graduate Migration Behavior</p></title>
  <aug>
    <au><snm>Faggian</snm><fnm>A</fnm></au>
    <au><snm>McCann</snm><fnm>P</fnm></au>
    <au><snm>Sheppard</snm><fnm>SC</fnm></au>
  </aug>
  <source>Journal of Regional Science</source>
  <pubdate>2007</pubdate>
  <volume>47</volume>
  <issue>3</issue>
  <fpage>517</fpage>
  <lpage>-539</lpage>
</bibl>

<bibl id="B7">
  <title><p>Topology of the Erasmus student mobility network</p></title>
  <aug>
    <au><snm>Derzsi</snm><fnm>A</fnm></au>
    <au><snm>Derzsy</snm><fnm>N</fnm></au>
    <au><snm>K{\'a}ptalan</snm><fnm>E</fnm></au>
    <au><snm>N{\'e}da</snm><fnm>Z</fnm></au>
  </aug>
  <source>Physica A: Statistical Mechanics and its Applications</source>
  <publisher>Elsevier</publisher>
  <pubdate>2011</pubdate>
  <volume>390</volume>
  <issue>13</issue>
  <fpage>2601</fpage>
  <lpage>-2610</lpage>
</bibl>

<bibl id="B8">
  <title><p>International Standard Classification of Education: ISCED
  2011</p></title>
  <aug>
    <au><cnm>UIS</cnm></au>
  </aug>
  <pubdate>2012</pubdate>
</bibl>

<bibl id="B9">
  <title><p>ISCED Fields of Education and Training 2013 (ISCED-F 2013): manual
  to accompany the International Standard Classification of
  Education</p></title>
  <aug>
    <au><cnm>UIS</cnm></au>
  </aug>
  <pubdate>2014</pubdate>
</bibl>

<bibl id="B10">
  <title><p>Student mobility in Higher Education. Erasmus and Erasmus
  Plus</p></title>
  <aug>
    <au><snm>Corradi</snm><fnm>S</fnm></au>
  </aug>
  <publisher>Rome: Laboratory of Lifelong Learning Department of Education and
  Training “Roma Tre” State University</publisher>
  <pubdate>2015</pubdate>
</bibl>

<bibl id="B11">
  <title><p>Erasmus+ Programme: Annual Report 2014</p></title>
  <aug>
    <au><snm>Commission</snm><fnm>E</fnm></au>
  </aug>
  <publisher>Publications Office of the European Union Luxembourg</publisher>
  <pubdate>2015</pubdate>
</bibl>

<bibl id="B12">
  <title><p>Investing in people. Making Erasmus even better</p></title>
  <aug>
    <au><snm>Commission</snm><fnm>E</fnm></au>
  </aug>
  <publisher>Publications Office of the European Union Luxembourg</publisher>
  <pubdate>2018</pubdate>
</bibl>

<bibl id="B13">
  <title><p>Three decades of Italian comparative advantages</p></title>
  <aug>
    <au><snm>De Benedictis</snm><fnm>L</fnm></au>
  </aug>
  <source>World Economy</source>
  <publisher>Wiley Online Library</publisher>
  <pubdate>2005</pubdate>
  <volume>28</volume>
  <issue>11</issue>
  <fpage>1679</fpage>
  <lpage>-1709</lpage>
</bibl>

<bibl id="B14">
  <title><p>Gender diversity in STEM disciplines: A multiple factor
  problem</p></title>
  <aug>
    <au><snm>Botella</snm><fnm>C</fnm></au>
    <au><snm>Rueda</snm><fnm>S</fnm></au>
    <au><snm>L{\'o}pez I{\~n}esta</snm><fnm>E</fnm></au>
    <au><snm>Marzal</snm><fnm>P</fnm></au>
  </aug>
  <source>Entropy</source>
  <publisher>Multidisciplinary Digital Publishing Institute</publisher>
  <pubdate>2019</pubdate>
  <volume>21</volume>
  <issue>1</issue>
  <fpage>30</fpage>
</bibl>

<bibl id="B15">
  <title><p>The Pursuit of Gender Equality-An Uphill Battle</p></title>
  <aug>
    <au><cnm>OECD</cnm></au>
  </aug>
  <pubdate>2017</pubdate>
</bibl>

<bibl id="B16">
  <title><p>Hierarchical structure and the prediction of missing links in
  networks</p></title>
  <aug>
    <au><snm>Clauset</snm><fnm>A</fnm></au>
    <au><snm>Moore</snm><fnm>C</fnm></au>
    <au><snm>Newman</snm><fnm>ME</fnm></au>
  </aug>
  <source>Nature</source>
  <publisher>Nature Publishing Group</publisher>
  <pubdate>2008</pubdate>
  <volume>453</volume>
  <issue>7191</issue>
  <fpage>98</fpage>
  <lpage>-101</lpage>
</bibl>

<bibl id="B17">
  <title><p>Heavy-tailed distributions for agent-based economic
  modelling</p></title>
  <aug>
    <au><snm>Clementi</snm><fnm>F</fnm></au>
  </aug>
  <source>Economics with Heterogeneous Interacting Agents</source>
  <publisher>Switzerland: Springer</publisher>
  <pubdate>2016</pubdate>
  <fpage>157</fpage>
  <lpage>-190</lpage>
</bibl>

<bibl id="B18">
  <title><p>Mean-field theory for scale-free random networks</p></title>
  <aug>
    <au><snm>Barab{\'a}si</snm><fnm>AL</fnm></au>
    <au><snm>Albert</snm><fnm>R</fnm></au>
    <au><snm>Jeong</snm><fnm>H</fnm></au>
  </aug>
  <source>Physica A: Statistical Mechanics and its Applications</source>
  <publisher>Elsevier</publisher>
  <pubdate>1999</pubdate>
  <volume>272</volume>
  <issue>1-2</issue>
  <fpage>173</fpage>
  <lpage>-187</lpage>
</bibl>

<bibl id="B19">
  <title><p>Fitting Heavy Tailed Distributions: The poweRlaw
  Package</p></title>
  <aug>
    <au><snm>Gillespie</snm><fnm>CS</fnm></au>
    <au><cnm>others</cnm></au>
  </aug>
  <source>Journal of Statistical Software</source>
  <publisher>Foundation for Open Access Statistics</publisher>
  <pubdate>2015</pubdate>
  <volume>64</volume>
  <issue>i02</issue>
</bibl>

<bibl id="B20">
  <title><p>Power-law distributions in empirical data</p></title>
  <aug>
    <au><snm>Clauset</snm><fnm>A</fnm></au>
    <au><snm>Shalizi</snm><fnm>CR</fnm></au>
    <au><snm>Newman</snm><fnm>ME</fnm></au>
  </aug>
  <source>SIAM review</source>
  <publisher>SIAM</publisher>
  <pubdate>2009</pubdate>
  <volume>51</volume>
  <issue>4</issue>
  <fpage>661</fpage>
  <lpage>-703</lpage>
</bibl>

<bibl id="B21">
  <title><p>Likelihood ratio tests for model selection and non-nested
  hypotheses</p></title>
  <aug>
    <au><snm>Vuong</snm><fnm>QH</fnm></au>
  </aug>
  <source>Econometrica: Journal of the Econometric Society</source>
  <publisher>JSTOR</publisher>
  <pubdate>1989</pubdate>
  <fpage>307</fpage>
  <lpage>-333</lpage>
</bibl>

</refgrp>
} 

\end{document}